\documentclass{cernrep}
\usepackage{texnames}
\usepackage[T1]{fontenc}
\usepackage[bookmarks, colorlinks=true, linktoc=page, pdftex, linkcolor=red, citecolor=red, urlcolor=red]{hyperref}
\RequirePackage{siunitx}
\begin{document}

\title{Laser-Driven Structure-Based Accelerators: White Paper for Snowmass 2021
Topical Group AF06 - Advanced Acceleration Concepts}
\author {R. J. England, D. Filippetto, G. Torrisi, A. Bacci, G. Della Valle, D. Mascali, G. S. Mauro, G. Sorbello, P. Musumeci, J. Scheuer, B. Cowan, L. Sch\"achter, Y-C. Huang, U. Niedermayer, W. D. Kimura, R. Li, R. Ischebeck, E.I. Simakov, P. Hommelhoff, R. L. Byer}

\maketitle 

\section{Executive Summary}
\label{sec:executive}
{\hskip 0.13in}

Significant progress has been made in the last 10 years in developing laser-driven microstructure accelerators based on dielectric laser acceleration (DLA), plasmonically-enhanced metasurface laser accelerators (MLA), and various photonic crystal configurations, which we here collectively refer to as laser-driven structure-based accelerators (LSA). As an advanced accelerator concept, this approach offers some unique advantages. The acceleration mechanism is inherently linear and occurs in a vacuum region in a static structure. In addition to the stability benefits this affords, the acceleration effect is inherently dependent on the phase of the laser field, which makes it possible to dynamically fine-tune accelerator performance by manipulation of the incident laser phase profile. The acceleration mechanism also works equally well for positrons and electrons. Due to its unique low-charge high repetition rate bunch format, this approach may provide a possible source technology for proposed fixed-target light dark matter searches using single few-GeV electrons. In addition, the projected beamsstrahlung energy loss for a multi-TeV collider scenario is in the few percent range, as opposed to tens of percents for conventional RF accelerators. Gradients on the GV/m scale have already been demonstrated with energy gains exceeding 0.3 MeV for a few-fC beam, and wall plug efficiencies comparable or superior to conventional approaches appear feasible. Furthermore, the primary supporting technologies (solid state lasers and nanofabrication) are mature and already at or near the capabilities required for a full-scale accelerator based on this approach. The LSA approach could significantly lower the cost per GeV by leveraging commercial developments of the integrated circuit and telecommunications industries. These advantages motivate supporting laser-driven structure accelerators as a competitive higher gradient alternative to more conventional RF accelerators. Development of near-term applications in medicine, science, and industry could utilize the unique capabilities of these sources while providing platforms for further technological development. The primary U.S. funding source for this area of research is a private grant from the Moore Foundation, which will expire in 2022. While synergistic applications (i.e., those needing ultra-compact accelerators) provide an opportunity for cost-sharing the technology development, overall the LSA technology has had significantly lower funding levels than other advanced accelerator technologies and essentially no funding dedicated to closing technology gaps unique for multi-TeV linear colliders. This is partly due to the perception that there is not a credible technology path for a 30 MW linear collider beam due to the small, micron-sized apertures, even when dividing the main beam power into multiple, parallel accelerating channels. This needs to be addressed as early in the LSA technology roadmap as possible. The key steps needed for a LSA linear collider development include the following near-term technology demonstrations: (1) low-cost high-efficiency dielectric structures with sufficient thermal conductivity and controllable wakefield effects; (2) focusing schemes with sufficiently high gradient that minimize beam interception but do not dilute the beam quality; and (3) low-cost, high-efficiency, and high-power drive lasers that have sub-cycle phase and timing control. Successful demonstration of these capabilities will show that the LSA technology is a credible alternative multi-TeV linear collider technology and provide motivation for a future multi-stage linear collider prototype.

\section{Introduction}
\label{sec:intro}
{\hskip 0.13in}
Constraints on the size and cost of accelerators have inspired a variety of advanced acceleration concepts for making smaller and more affordable particle accelerators. The use of lasers as an acceleration mechanism is particularly attractive due to the intense electric fields they can generate combined with the fact that the solid state laser market has been driven by extensive industrial and university use toward higher power and lower cost over the last 20 years. Dielectrics and semiconductor materials have optical damage limits corresponding to acceleration fields in the 1 to 10 GV/m range, which is orders of magnitude larger than conventional accelerators.  Such materials are also amenable to rapid and inexpensive CMOS and MEMS fabrication methods developed by the integrated circuit industry.  These technological developments over the last two decades, combined with new concepts for efficient field confinement using optical waveguides and photonic crystals, and the first demonstration experiments of near-field structure-based laser acceleration conducted within the last few years, have set the stage for making integrated laser-driven  accelerators for a variety of applications. We here collectively refer to schemes of this type as laser-driven structure-based accelerators (LSA), which includes concepts employing all-dielectric materials ("dielectric laser accelerators" or DLA), plasmonically enhanced "metasurface laser accelerators" (MLA), and variants using photonic crystals of various 1D, 2D, and 3D geometries. 

\begin{figure}
\begin{center}
 \includegraphics[height=.35\textheight]{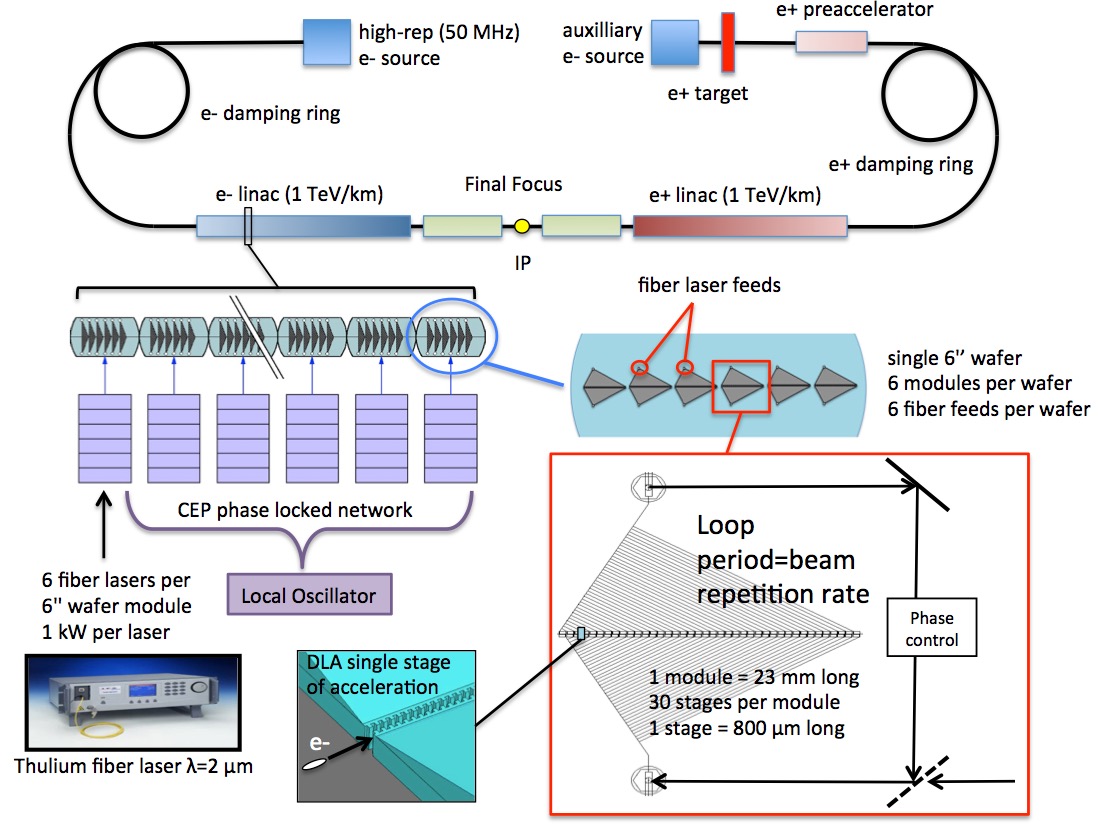}
\caption{Conceptual schematic of a multi-TeV e+ e- collider driven by a carrier envelope phase locked network of energy-efficient solid-state fiber lasers at 20 MHz repetition rate. Laser power is distributed by photonic waveguides to a sequence of dielectric accelerating, focusing, and steering elements co-fabricated on 6-inch wafers which are aligned and stabilized using mechanical and thermal active feedback systems.}
\label{collider}
\end{center}
 \end{figure}

A future laser-driven linear collider, schematically illustrated in Fig. \ref{collider}, will require the development of high-gradient accelerator structures as well as suitable diagnostics and beam manipulation techniques, including compatible small-footprint deflectors, focusing elements, and beam position monitors (BPMs).  Key developments in these areas have been made in recent years, including the demonstration of high average gradients (300--850 MeV/m) at relativistic particle energies \cite{peralta:2013, wootton_demonstration_2016, cesar_nonlinear_2018}, non-relativistic acceleration with gradients up to 350 MV/m \cite{breuer_laser-based_2013,leedle_dielectric_2015}, and development of preliminary design concepts for compatible photonic components and power distribution networks \cite{hughes:chip:2018,mcneur_elements_2018}. The power distribution scheme is then envisioned as a fiber-to-chip coupler that brings a pulse from an external fiber laser onto the integrated chip, distributes it between multiple structures via on-chip waveguide power splitters, and then recombines the spent laser pulse and extracts it from the chip via a mirror-image fiber output coupler \cite{colby:2011}, after which the power is either dumped, or for optimal efficiency, recycled \cite{siemann:2004}.  Maintaining phase synchronicity of the laser pulse and the accelerated electrons between many separately fed structures could be accomplished by fabricating the requisite phase delays into the lengths of the waveguide feeds and employing the use of active feedback systems. The LSA mechanism is also equally suitable for accelerating both electrons and positrons. Example machine parameters have been outlined in the Snowmass 2013 report and several other references \cite{colby:2011,dla:2011,snowmass:2013,england:rmp2014}.  In these example parameter studies, desired luminosities appear feasible with reasonable power consumption and with low beamstrahlung energy loss \cite{beambeam:2021}.  

\section{Background}
\label{sec:background}
{\hskip 0.13in}

\subsection{Accelerator Structures}

Use of lasers to accelerate charged particles in material structures has been a topic of considerable interest since shortly after the optical laser was invented in the early 1960s.  Early concepts proposed using lasers to accelerate particles by operating known radiative processes in reverse, including the inverse Cherenkov accelerator \cite{shimoda:1962} and the inverse Smith-Purcell accelerator \cite{takeda:1968, palmer:1980}.  Energy modulation of relativistic electrons has also been observed in a laser field truncated by a thin downstream metallic film \cite{leap:2005,sears:2008}.  In these direct optical-scale interactions, an electron bunch longer than the operating wavelength of the accelerator will sample all phases of the accelerating field and will therefore experience an energy modulation.  In order to produce a net acceleration of the electrons using the DLA concept, the bunch must therefore be microbunched with a periodicity equal to the laser wavelength.  Techniques for accomplishing this at the optical period of a laser have been previously demonstrated at wavelengths of 10 $\mu$m and 800 nm \cite{kimura:2001,sears:atto2008}. 

\begin{figure}
\begin{center}
 \includegraphics[width=0.9\textwidth]{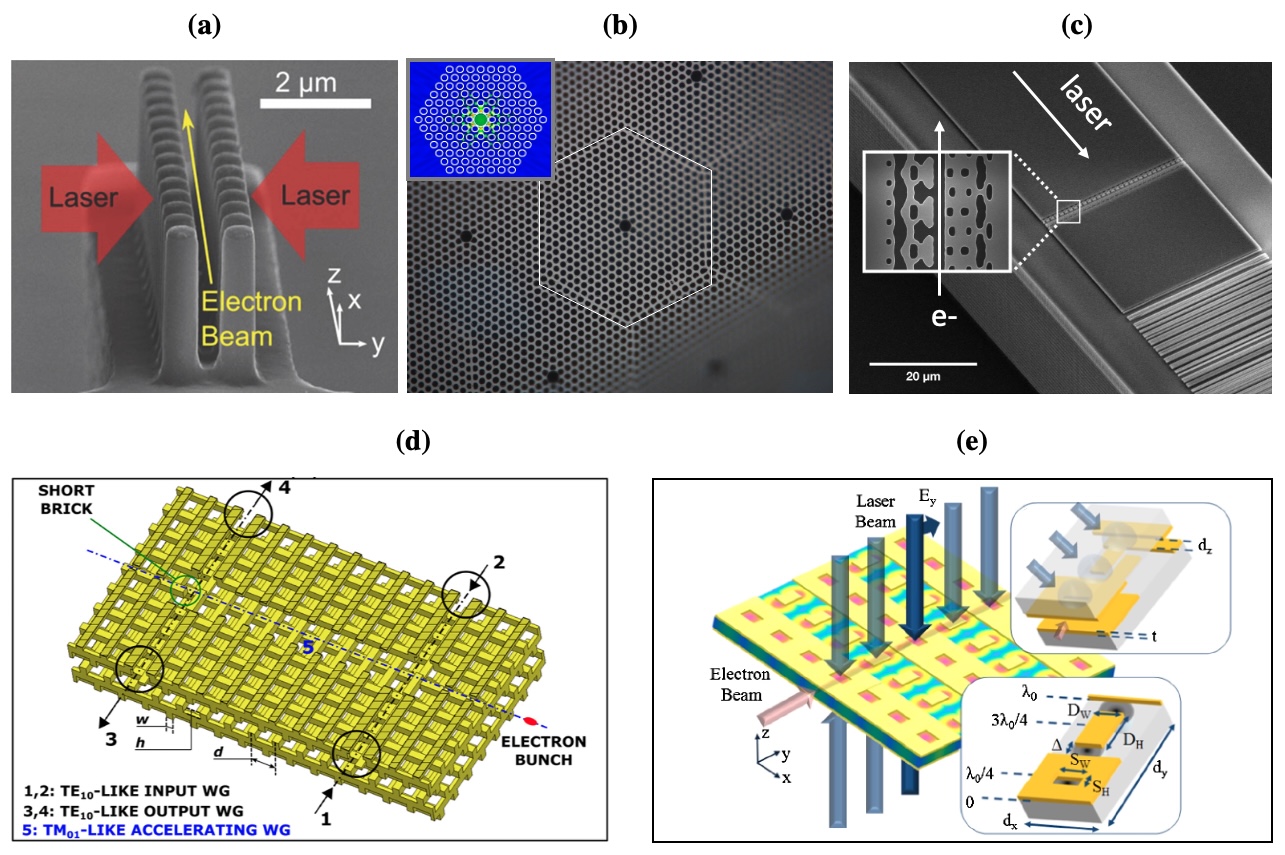}
\caption{Various LSA structures: (a) the planar-symmetric dual-pillar DLA geometry \cite{leedle:2018}; (b) a hexagonal hollow-core photonic crystal fiber with TM$_{01}$ like mode (inset) \cite{noble:2011};  (c) a waveguide-coupled inverse designed accelerator \cite{sapra:science2020}; (d) woodpile structure \cite{mauro2021woodpile} slice; (e) example of MLA structure \cite{bar-lev_plasmonic_2014} }
\label{structures}
\end{center}
 \end{figure}

These initial experiments laid the ground work for efficient phased laser acceleration.  However the interaction mechanism used to accelerate the particles in the experiments of Refs. \cite{takeda:1968, palmer:1980,leap:2005,sears:2008} is a relatively weak effect, requiring laser operation at fluences above the damage limit of the metal surface.  This points to the need to use materials with high damage limits combined with acceleration mechanisms that are more efficient.  Due to these considerations, a shift in focus has occurred towards photonic structures made of dielectric materials and incorporating new technologies such as photonic crystals and meta-surfaces \cite{rosing:1990,lin:2001,cowan:2003,mizrahi:2004,schachter:2004,naranjo:2012,bar-lev_plasmonic_2014}.  The slab-symmetric or planar 1D type of geometry illustrated in Fig. \ref{structures}(a) is simpler to fabricate and its wide aspect ratio helps in improving charge transmission, making demonstration experiments simpler.  This has led to structures of type (a) being the first to be successfully fabricated and undergo demonstration experiments \cite{peralta:2013,breuer:2013,leedle:2015}. 

In addition to the 1D approach of Fig.  \ref{structures}(a) there are several attempts closer to conventional RF accelerator where a TM or TM-like mode is guided by the interaction structure, as in Figs. \ref{structures}(b),(d). The use of inverse design codes additionally allows for sophisticated optimization, such as the first demonstrated waveguide-coupled microchip accelerator shown in Fig. \ref{structures}(c) from Ref. \cite{sapra:science2020}. Among the hollow-core dielectric Photonic Crystal (PhC) structures suitable for DLAs, the woodpile configuration has a fully 3D frequency band-gap \cite{hermatschweiler2007fabrication, staude2012waveguides}. An example of Electromagnetic Bandgap (EBG) hollow-core woodpile waveguide,  with an appropriate side-coupler (or mode launcher-converter) \cite{mauro2021woodpile} is visible in Fig.~\ref{structures} (d).  The hollow core structures are well suited to handle high power with low loss \cite{michieletto2016hollow}.  
When the wave reaches the end of the accelerating channel (that can be ``arbitrarily'' long), it is back-converted into two waves which are picked-up by a second mode converter (3 \& 4) and driven out of the structure. 
The travelling wave, picked-up at the accelerator output, can be absorbed without reflection on a load or amplified and re-injected at the input port \cite{siemann2004energy,na2005energy}.

Another class of structure, which we refer to as a \textit{metasurface laser accelerator} (MLA), takes advantage of the near field enhancement generated by nano-structured plasmonic materials ~\cite{economou_surface_1969}. As illustrated in Fig. \ref{structures}(e), a laser field impinging on such a surface can excite traveling waves confined at the metal-dielectric interface, called surface plasmon polaritons (SPP), leading to large local field enhancement. Realization of SPP nano-cavities for electron generation and acceleration has been experimentally demonstrated to generate very large field enhancements~\cite{polyakov_plasmonic_2011}. The nanostructured surface can be precisely engineered to obtain the amplitude and phase profile required for optimized acceleration of electron pulses \cite{bar-lev_plasmonic_2014,bar-lev_design_2019}. 
Although metals sustain higher ohmic losses with respect to dielectric systems, the large field enhancement obtained relaxes the requirements on the incident laser intensity for the same accelerating gradient. In addition, the fabrication and patterning of thin metallic films is significantly simpler than that of high aspect ratio dielectric structures. While MLA structures can achieve very high field enhancements, the power density stored locally can quickly generate damage, especially at high repetition rates, requiring the use of non-resonant structures \cite{bar-lev_plasmonic_2014,bar-lev_design_2019}.  

Most of the LSA structures proposed to date, including those in Fig. \ref{structures} take advantage of intrinsically low quality-factor Q of the geometry, which
results in lower field enhancement, but also decreases the energy density stored locally and
therefore the potential for structure damage. In addition, the lower Q-factor supports wider optical bandwidth, facilitating the use of ultra-fast laser sources. Due to the stringent requirements of a linear collider on beam quality and luminosity, we outline below various key constraints on structure design and beam operation which we will then observe in developing a strawman collider scenario in Section \ref{sec:parameters}. 


\subsection{Current State of the Art}
The field of LSAs has undergone significant experimental progress within the last few years. In 2013, the first demonstrations of acceleration in a DLA were performed by teams at SLAC and at Friedrich Alexander University \cite{peralta:2013,breuer:2013}.  In the former experiment, a dual grating structure was illuminated by laser pulses from a Ti:Sapphire laser as the 60 MeV electrons from the Next Linear Collider Test Accelerator traversed the vacuum channel between the fused silica gratings. Acceleration with gradients approaching 300 MeV/m was observed. In the latter experiment, the optical near fields of a fused silica single grating, illuminated by laser pulses from a Ti:Sapphire oscillator cavity, accelerated 28 keV electrons with a gradient of 25 MeV/m. Since these proof of principle experiments, different structures, materials, and lasers have been used to improve the acceleration gradients. Additionally, other features of the accelerating fields (e.g. their sub-optical cycle structure and their deflecting forces) have been probed. At Stanford, silicon dual pillar structures, illuminated by a Ytterbium fiber laser, were used to accelerate 100 keV electrons with an acceleration gradient of 370 MeV/m \cite{leedle:2015}. Further, in this experiment, the transverse deflecting fields of the accelerating mode were characterized. To increase the accelerating gradient further, short pulsed lasers have recently been used. By utilizing a 100 fs pulse length Ti:Sapphire laser, a group at SLAC and UCLA were able to demonstrate accelerating gradients of up to 850 MeV/m for dual grating fused silica DLAs and 8 MeV electrons \cite{cesar_nonlinear_2018}. Similarly, by using a 20 fs pulse length laser, a group at FAU Erlangen was able to accelerate 28 keV electrons with a gradient of 210 MeV/m \cite{kozak_dielectric_2017}. The transverse control of 28 keV electrons has recently been demonstrated with both transverse deflecting structures (a rotated silicon grating) and focusing structures (a Si grating with parabolically shaped grating teeth) \cite{kozak_dielectric_2017}. The ultrafast structure of the fields has led to ultrafast gating of electrons with applications in ultrafast electron diffraction (UED).  The rapid pace of recent progress in this field sets the stage for developing LSA for a variety of applications. However, moving from proof of principle demonstrations towards a laser-driven electron-positron collider poses significant technical challenges.

\subsection{Near Term Applications}

With tabletop sources now coming into operation in university labs, near-term applications that utilize presently available low-current beams with moderate particle energies in the 100 keV to few MeV range are being actively pursued. Due to the intrinsic optical-scale bunch structure, with sub-femtosecond bunch duration, compact electron sources for ultrafast science and electron diffraction studies are among the most promising applications. Compact accelerators with target energies in the few MeV range for medical dosimetry also provide a compelling near-term use for LSA technology. The sub-femtosecond high brightness electron sources based on laser-driven on-chip accelerators could potentially be fit on the end of an optical fiber, placed on a scanning platform at the surface of a sample, or even inserted into living tissue, to supply unprecedented time, space, and species resolution.

\begin{figure}
\begin{center}
 \includegraphics[height=.18\textheight]{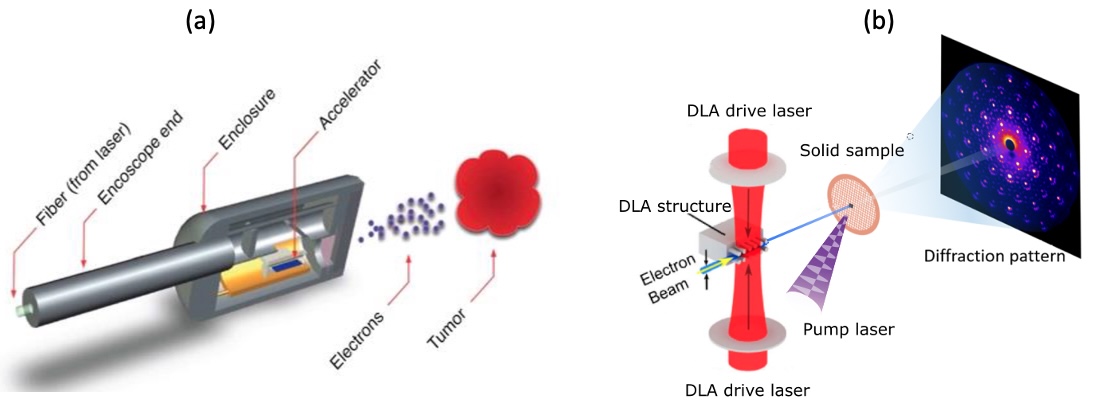}
\caption{Promising near-term applications for laser-driven accelerators include (a) endoscopic medical radiation therapy delivery devices, (b) tabletop sources for MeV ultrafast electron diffraction studies.}
\label{applications}
\end{center}
 \end{figure}

An ultracompact self-contained multi-MeV electron source based on integrated photonic particle accelerators could enable minimally invasive cancer treatments and adjustable dose deposition in real-time, with improved dose control. For example, one could envision an encapsulated micro-accelerator built onto the end of a fiber-optic catheter placed within a tumor site using standard endoscopic methods, allowing a doctor to deliver the same or higher radiation dose to what is provided by existing external beam technologies, with less damage to surrounding tissue \cite{england:aac14, DOE:CASM:2020}. As electrons have an energy loss rate of about 2 MeV/cm in water, their irradiation volumes can be tightly controlled. Encapsulated devices would ideally have variable electron energies in the 1-10 MeV range, a footprint that is millimeter-scale, and accommodate a wide range of emission angles for various treatment modalities. The manufacturing and operating costs based on low-cost or disposable chips, powered by an external fiber laser, could be much lower than those for conventional radiation therapy machines, and the robustness of such systems compared to conventional accelerators could be even more favorable.

Pulse trains of attosecond electron bunches are intrinsic to the LSA approach and could provide excellent probes of transient molecular electronic structure. Recently, this fine time-scale structure has been experimentally measured (with bunch durations from 270 to 700 attoseconds) and injected into a subsequent acceleration stage to perform fully on-chip bunching and net acceleration demonstrations \cite{black:Atto:2019,schoenenberger:Atto:2019}. At present, attosecond electron probes are limited to strong-field experiments where laser field-ionized electrons re-scatter from their parent ion. The far greater flexibility of on-chip attosecond electron sources could be important complements to optical attosecond probes based on high harmonics or attosecond X-ray free electron lasers (FELs). The capabilities of chip-scale electron probes could be transformational for studies of chemical impurities and dopants in materials. Many nanoscale technologies operate by inter- and intra-molecular processes with intrinsic timescales measured in femtoseconds; proton migration in materials occurs over tens of femtoseconds, and electron migration over single femtoseconds. Examples include quantum memory storage and switching using atomic impurities in pure crystalline hosts and lab-on-a-chip technologies for chip-scale genome decoding. This technology could also enable a unique class of compact tabletop electron sources for ultrafast electron microscopy (UEM). Since both the accelerator and associated drive laser and peripherals are all compact enough to fit on a tabletop, this would enable the construction of compact and flexible attosecond electron and photon sources that can resolve atomic vibrational states in crystalline solids in both spatial and temporal domains simultaneously. This could lead to improved understanding of the dynamics of chemical reactions at the level of individual molecules, the dynamics of condensed matter systems, and photonic control of collective behaviors and emergent phenomena in quantum systems.

The difference in bunch charge and duration for optically accelerated electron beams also points to the potential for future light sources for generation of attosecond-scale pulses of extreme ultraviolet (EUV) or X-ray radiation, with the potential to produce extremely bright electron beams that are suitable for driving superradiant EUV light in a similarly optical-scale laser-induced undulator field \cite{england:FLS2018}. Furthermore, because of the few-femtosecond optical cycle of near infrared mode-locked lasers, laser-driven free electron lasers (FELs) could potentially generate attosecond X-ray pulses to probe matter on even shorter time scales than is possible today. Laser-driven dielectric undulators have been proposed and could be fabricated using similar photolithographic methods used to make on-chip accelerators \cite{plettner:2008,plettner:2009}. Combining the high gradient and high brightness of advanced accelerators with novel undulator designs could enable laboratory-scale demonstrations of key concepts needed for future EUV and X-ray lasers that hold the potential to transform the landscape of ultrasmall and ultrafast sciences. To realize these laboratory-scale, lower-cost, higher performance radiation sources, critical components of laser-driven FELs need to be developed and demonstrated. Coherent attosecond radiation could potentially be produced using the same operating principles that produce particle acceleration via the “accelerator on a chip" mechanism. These structures operate optimally with optical-scale pulse formats, making megahertz repetition rate attosecond-scale pulses a natural combination. The optical-scale FEL regime has not been extensively studied before and questions arise as to how well the beam will behave in such structures and how well it will ultimately perform. The theoretical and numerical tools to model these processes need to be developed in order to guide experimental studies of attosecond electron and photon generation.

Various technical challenges to developing near-term applications for tabletop laser accelerators (including high-energy gain, multi-stage beam transport, and higher average beam power) must also be addressed in order to reach the stringent beam quality and machine requirements for a HEP collider. The near-term applications outlined above therefore represent an intermediate advancement of the technology towards ultimately desired HEP performance levels. Further development of these applications also holds the potential to leverage both scientific and industrial interests that could facilitate more rapid development and support. 

\subsection{Intermediate Term Applications}



The nature and the origin of dark matter are arguably among the biggest mysteries in physics. Dark matter particles that have a mass smaller than the proton, termed \textit{light dark matter}, are among the candidates to solve this mystery. It has been recently proposed to search for light dark matter by observing missing energy and transverse momentum during the interaction of individual electrons with a fixed target at few-GeV particle energies \cite{LDMX}. To this aim, it is important to prepare an initial state with precisely known parameters, and to measure deviations from standard model predictions with the highest possible accuracy. Due to the lack of electromagnetic and strong interactions, the cross section for the production of light dark matter is extremely small. A high repetition rate of the particle source and a fast readout of the detector are thus important. Even with an interaction rate on the order of GHz, it is expected that several years of data acquisition will be required to find a signal of dark matter events. Generating a beam of single electrons appears possible with radio frequency accelerators \cite{eSPS}, but it is difficult to do so with high efficiency. The required electron energies in the GeV range are also out of reach of DC sources. Dielectric laser accelerators have been discussed as possible high-repetition-rate sources of single electrons that could help fill this source technology requirement. Although generating beams for applications in dark matter search is beyond present capabilities of DLA technology, such an application could be envisioned as an intermediate step on the road towards an accelerator at the energy frontier.

\section{Collider Concept}
\label{sec:concept}
{\hskip 0.13in}

\subsection{Strawman Parameters}
\label{sec:parameters}
{\hskip 0.13in}
To reach 30 TeV center-of-mass energies, a next generation lepton collider based on traditional RF microwave technology would need to be over 100 km in length and would likely cost tens of billions of dollars to build.  Due to the inverse scaling of the interaction cross section with energy, the required luminosity for such a machine would be as much as $100\times$ higher than proposed 1 to 3 TeV machines (ILC and CLIC), producing a luminosity goal of order $10^{36}$ cm$^{-2}$s$^{-1}$.  In attempting to meet these requirements in a smaller cost/size footprint using advanced acceleration schemes, the increased beam energy spread from radiative loss during beam-beam interaction (beamstrahlung) at the interaction point becomes a pressing concern.  Since the beamstrahlung parameter is proportional to bunch charge, a straightforward approach to reducing it is to use small bunch charges, with the resulting quadratic decrease in luminosity compensated by higher repetition rates.  This is the natural operating regime of the LSA scheme, with the requisite average laser power (>100 MW) and high (>10 MHz) repetition rates to be provided by modern fiber lasers.

Strawman parameters for the 250 GeV and 3 TeV cases have been previously reported \cite{england:rmp2014,rast:2016}, and these are reproduced in Table \ref{parameters}.  To scale this scenario to 30 TeV we note that the total wall-plug power is proportional to the beam power $P_\text{wall} = P_\text{beam}/\eta$, where $\eta$ is the wall-plug efficiency and the beam power (of both beams together) is $P_\text{beam} = E_\text{cm} n N f_\text{rep}$, where $n$ is the number of bunches per train, $N$ the number of electrons per optical microbunch and $E_\text{cm}$ the center-of-mass energy.  The geometric luminosity scales as
\begin{equation}
\mathcal{L} = \frac { (n N)^2 f_\text{rep}} {4 \pi \sigma_x \sigma_y} = \chi E_\text{cm}^2 ,
\label{eq:luminosity}
\end{equation}
where $\chi = 2.3 \times 10^{33} \text{cm}^{-2} \text{s}^{-1} \text{TeV}^{-2}$ is a scaling constant \cite{king:2000}. We note that the luminosty here scales as $(n N)^2$ rather than $n N^2$ because it is assumed that entire bunch trains (each a single laser pulse in duration) collide at the IP.  For purposes of calculating the disruption parameter, luminosity enhancement, and beamstrahlung energy loss, it is further assumed that the microbunch structure is smeared out prior to the IP, giving a luminosity enhancement of approximatey 10.  Hence the required particle flux scales as 
\begin{equation}
n N f_\text{rep} = E_\text{cm} \sqrt{4 \pi \sigma_x \sigma_y \chi f_\text{rep}} .
\label{eq:flux}
\end{equation}
Combining these relations we obtain the following scaling for wall-plug power with center-of-mass energy:
\begin{equation}
P_\text{wall} = \eta^{-1} E_\text{cm}^2 \sqrt{ 4 \pi \sigma_x \sigma_y \chi f_\text{rep}} .
\label{eq:wallplug}
\end{equation}
Consequently, if the center of mass energy is increased by a factor of 10 (from 3 TeV to 30 TeV), then for similar repetition rate and efficiency, the wall-plug power will increase by a factor of 100, as reflected in Table \ref{parameters}.  At the same time, Eq.~\eqref{eq:flux} requires a 10 times increase in average beam current.  Due to the scaling in Eq.~\eqref{eq:flux} with $f_\text{rep}$ there is a tradeoff between charge and repetition rate. However, since the bunch charge $N$ and laser pulse duration are constrained by the efficiency, gradient, and space charge arguments of Section \ref{structures}, a potential solution is to incorporate 10 parallel beamlines in a matrixed configuration such as that of the 2D honeycomb structure of Fig.~\ref{structures}(b) or as described in Ref. \cite{mimosa:2020}. The physics of the beam recombination mechanism at the IP requires further study, but for the purposes of Table \ref{parameters} we assume a linear emittance scaling with number of parallel beamlines. In these examples, desired luminosity appears feasible and with a small (few percent) beamstrahlung energy loss.  Although the numbers in Table \ref{parameters} are merely projections used for illustrative purposes, they highlight the fact that due to its unique operating regime, laser-driven accelerators are poised as a promising technology for future collider applications.

\begin{table}[]
\caption{Strawman Parameters at 250 GeV, 3 TeV, and 30 TeV Energies}
\label{parameters}
\begin{center}
\begin{tabular}{lllll}
\hline
Parameter                 & Units & 250 GeV & 3 TeV   & 30 TeV  \\ \hline
Center of Mass Energy     & GeV   & 250     & 3000    & 30000   \\
Bunch charge              & e     & 3.8e4   & 3.0e4   & 3.0e4   \\ 
\# Bunches/train          & \#    & 159     & 159     & 159     \\ 
\# Parallel Beamlines     & \#    & 1       & 1       & 10      \\
Train Repetition Rate     & MHz   & 20      & 20      & 20      \\
Final Bunch Train Length  & ps    & 1.06    & 1.06    & 1.06    \\
Single Bunch Length       & $\mu$m    & 2.8e-3   & 2.8e-3   & 2.8e-3   \\
Drive Wavelength          & $\mu$m    & 2       & 2       & 2       \\
IP X Emittance (Norm)     & nm    & 0.1     & 0.1     & 1       \\
IP Y Emittance (Norm)     & nm    & 0.1     & 0.1     & 1       \\
IP X Spot Size            & nm    & 2       & 1       & 1       \\
IP Y Spot Size            & nm    & 2       & 1       & 1       \\
Beamstrahlung Energy Loss & \%    & 0.6     & 1.0     & 2.6     \\
Length of Beam Delivery   & m     & 2321    & 2304    & 2304    \\
Effective L*              & m     & 5       & 5       & 5       \\
Total Length              & km    & 5.0     & 8.4     & 42.1    \\
Geometric Luminosity      & cm$^{-2}$s$^{-1}$  & 1.46e33 & 3.63e33 & 3.63e35 \\
Enhanced Luminosity       & cm$^{-2}$s$^{-1}$  & 1.84e34 & 3.19e34 & 3.19e36 \\
Beam Power (per beam)     & MW    & 2.4     & 22.9    & 2292.5  \\
Total Wallplug Power      & MW    & 88.1    & 360.3   & 30487.4 \\
Wallplug Efficiency       & \%    & 5.5     & 12.7    & 15.0    \\ \hline
\end{tabular}
\end{center}
\end{table}

\subsection{Technological Requirements}
\label{sec:requirements}
{\hskip 0.13in}

\textbf{Particle Sources}. Considerable effort at the university level has been directed towards development of compact electron sources based on laser-assisted field emission from nanotips, which can produce electron beams with very small emittance values. While such miniaturized low-current particle sources are well adapted to the LSA approach and are highly desirable for a variety of applications, they are not well suited to the required beam power of a collider facility. High gradient, high energy superconducting radio-frequency (SRF) guns operated in continuous wave (CW) mode are promising candidates for delivering relevant beams for a linear collider. This scenario is illustrated in Fig. \ref{collider}, where SRF guns are proposed as a source for electron generation, and positron generation via a target. The CW operation allows in principle that each RF bucket can be filled with a photoelectron bunch up to the resonant frequency of the cavity, typically 1.3 GHz for elliptical geometry guns and 100-200 MHz for quarter-wave resonator type guns \cite{arnold:2011}. Preliminary simulations show that it is possible to deliver 10 fC, 1 ps, 1.0 nm-rad emittance, 2 MeV electron beams from a 2$\mu$m RMS laser spot on the cathode with 0.2 mm-mrad/mm RMS intrinsic emittance in a 20 MV/m, 200 MHz quarter-wave resonator type SRF gun. The clean vacuum environment inside SRF guns also potentially allows for advanced photocathodes to be used, while care must be taken to avoid contamination of the SRF cavity surface by nanoparticles from the cathode. To meet the brightness requirements for injection into LSA structures, a local enhancement of electric field can be achieved through surface plasmon polariton (SPP) interference. This approach allows for a sufficiently broadband (i.e. ultrafast) response \cite{durham_plasmonic_2019}. In addition, the highly confined multiphoton photoemission from a flat surface avoids aberrations and damage issues intrinsic to curved or tip-like photocathodes in high field environments. A feasibility study is needed to understand how to adapt such a source to meet linear collider luminosity requirements.

\textbf{Emittance Damping.} Damping rings can be used to reduce the emittance below the intrinsic value established by the cathode, as illustrated in Fig. \ref{collider}. A damping ring operates by forcing particles to emit bremsstrahlung at the expense of all three momentum components, while compensating for the kinetic energy loss in the longitudinal direction. In this way, the transverse phase-space can in theory be suppressed to the quantum limit set by the period of the lattice. However, designing a damping ring compatible with optical scale bunch structure is a nontrivial engineering task. An alternative and potentially more compact method has been recently suggested, based on the use of an azimuthally symmetric optical Bessel beam to transversely focus the particles \cite{schachter:kimura:2020}. As shown in Ref. \cite{beambeam:2021}, when the radiation reaction force (RRF) is included, angular momentum is no longer conserved under strong focusing and transverse emittance can be suppressed. Using a causal version of the Abraham-Lorentz-Dirac equations, it can be shown theoretically that energy reduction of more than 35\% due to RRF can produce a similar reduction in the transverse emittance over a 1-meter interaction \cite{Schachter:OBB}. Consequently, a more compact alternative to a damping ring for laser accelerated particles could be envisioned by staging several such sections interspersed with compensating acceleration sections. 

\textbf{Single Mode Operation.}  The 2D (fiber) type geometry of Fig.~\ref{structures}(b) is similar conceptually to more conventional RF accelerators in that the confined accelerating mode is close to azimuthally symmetric, mimicking the transverse magnetic TM$_{01}$ mode of a conventional accelerating cavity.  However, the preferred fabrication technique (telecom fiber drawing) is less amenable to a fully on-chip approach. A lithographically produced variant of such an azimuthal structure, shown in Fig.~\ref{structures}(c), would allow confinement of a pure TM$_{01}$ accelerating mode, which is beneficial for stable beam transport over multi-meter distances. For such a structure, the radius of the vacuum channel is roughly $R = \lambda / 2$ whereas the bunch radius should satisfy $\sigma_r < 0.1 \lambda$, where $\lambda$ is the laser wavelength. This leads to two important aspects that should be emphasized. One is that due to constructive interference of a train of microbunches, the projection of the total wake on the fundamental mode is magnified. The other is that dielectric materials are virtually transparent over a large range of wavelengths. Consequently, whereas tens of thousands of modes are used for wake calculations in metallic RF cavities, only a few hundred modes contribute to long-range wake effects. 


\textbf{Guiding and Focusing.} For long-distance particle transport there is a serious need for a focusing system. Techniques for DLA have been proposed that utilize the laser field itself to produce a ponderomotive focusing force either by excitation of additional harmonic modes or by introducing drifts that alternate the laser field between accelerating and focusing phases to simultaneously provide acceleration as well as longitudinal and transverse confinement \cite{naranjo:2012,niedermayer:focusing:2018}. In simulation, such focusing techniques can adequately confine a particle beam to a narrow channel and overcome the resonant defocusing of the accelerating field. New structure designs and experiments are currently underway to test these approaches. It is also possible to use Spatial Harmonic Focusing~\cite{naranjo_stable_2012} or Alternating Phase Focusing (APF)~\cite{niedermayer:focusing:2018} to satisfy external focusing requirements. However, a generalization to 3D of the originally proposed two-dimensional schemes is required. As discussed in~\cite{Niedermayer_PRL_2020}, the 3D scheme has advantages at low energy, where the confinement to the extremely small aperture can also be provided in the vertical axis. Recently, based on this scheme and using Silicon-on-Insulator (SOI) wafers, a completely scalable multi-stage accelerator could be designed~\cite{Niedermayer_PhysRevApplied_2021}. At high energy, the 3D APF scheme allows stronger focusing gradients, since the square-sum of the two focusing constants scales with $\gamma^{-2}$~\cite{Niedermayer_PRL_2020}, such that in a counter-phase arrangement the two transverse planes can  exhibit focusing gradients that are not constrained by the beam energy. This allows for structure designs employing a single high-damage-threshold material, allowing for shorter focusing periods and higher emittances than the earlier 2D structures~\cite{Niedermayer_etal_this_issue}.

\subsection{Power Requirements and Energy Efficiency}
{\hskip 0.13in}
Assuming an average loaded gradient of 1 GV/m, the active length of each arm of the collider is 1.5 km yielding approximately 0.1 MW/m of average laser beam power per accelerating channel, with 33 lasers per meter of active accelerator length and 1 kW of average power required per laser. While CW lasers exceed 50\% efficiency at slightly lower power levels, high-repetition rate lasers currently cannot deliver the necessary average power.  However, laser experts predict that modern fiber lasers will reach the requisite average power levels within 5--10 years \cite{dla:2011}.  Close concentration of such laser energy in a dielectric substrate raises concern about heat dissipation. Compared with metallic surfaces at RF frequencies, the absorption coefficients for dielectrics at optical wavelengths are relatively low. It is found in Ref. \cite{karagodsky:2006} that, ignoring wake field effects, the heat dissipated by the fundamental mode in a Bragg waveguide is at least three orders of magnitude below the practical limit for thermal heat dissipation from planar surfaces (1500 W/cm$^2$). 

For the case of a 30 TeV collider, the average power carried by the electron beam is of order 0.5 GW. In order to get this power into the electron beam, we need twice this power (assuming 50\% laser-to-electron coupling efficiency) in the laser beam. It has been shown \cite{hanuka:single:2018} that maximum efficiency does not occur for the same parameters as maximum loaded gradient. Therefore we must either operate at maximum efficiency and compromise the gradient (increasing the required length of the accelerator) or operate at the maximum loaded gradient and compromise the efficiency and thus running into severe problems of wall-plug power consumption. Several possible regimes of operation have been analyzed in Ref. \cite{hanuka:regimes:2018}.  For our considerations in Section \ref{sec:parameters}, we take as a conservative number a loaded gradient of 1 GeV/m, and a multi-bunch laser-electron coupling efficiency of 40\%. By microbunching the beam, the coupling efficiency of the axial laser field to the particles in a DLA can in principle be as high as 60\% \cite{siemann:2004,na:2005}.  Combined with recent advances in power efficiencies of solid state lasers, which now exceed 30\% \cite{moulton:2009} and designs for near 100\% power coupling of laser power into a DLA structure \cite{wu:2014}, estimates of wall-plug power efficiency for a DLA based system are in the range of 10--12\%, which is comparable to more conventional approaches \cite{england:rmp2014}.

\subsection{Design Challenges and Performance Limits}
{\hskip 0.13in}

\textbf{Bunch Format.} Current laser technology permits repetition rates as high as 100 MHz with pulse durations on the order of picoseconds. Consequently, each such laser pulse may contain an electron train of 100 to 1000 microbunches.  The bunch structure and the aperture of the acceleration structure determine the number of electrons to be contained in one microbunch. The effective radial field $E_\perp = (e n \sigma_r / 2 \epsilon_0 \gamma^3)$ generated by a pencil beam is suppressed by the relativistic Lorentz factor $\gamma$, but becomes very significant at low energies. For example, in the non-relativistic case $10^4$ electrons confined in a sphere of radius 0.1 $\mu$m  is of the order of 1 GV/m, which is comparable with the laser field. Consequently, one micro-bunch can not contain more than on the order of a few $10^4$ electrons.  In the example parameters of Table \ref{parameters} we assume a microbunch charge of $3 \times 10^4$ electrons/positrons, which is also consistent with efficient multi-bunch operation.  

\textbf{Beam Break-Up (BBU)}.  According to Panofsky \cite{panofsky:BBU:1968}, BBU was first observed in 1966 as the pulse length of the transmitted beam appeared to shorten, provided the beam current exceeds a threshold value at a given distance along the accelerator; the greater the distance, the lower the threshold. Its essentials were found to be transverse fields generated by the beam \cite{chao:1980}. During the years BBU attracted attention every time a new acceleration paradigm came to serious consideration: this was the case for NLC \cite{dehler:1998} and CLIC \cite{braun:2008}, where it has been suggested to damp and detune the structure (DDS) in order to suppress hybrid high order modes (HOM) that lead to BBU instability. Later when the energy recovery linac (ERL) was in focus BBU was investigated in this configuration \cite{hoffstaetter:2004} and further for superconducting RF gun \cite{volkov:2011}. Without exception, the acceleration structure is initially azimuthally symmetric and the hybrid modes are excited due to transverse offset or asymmetry of the beam or to the coupling of input or output arms. In the case of dielectric structures, with the exception of Bragg waveguide \cite{mizrahi:2004}, the acceleration modes are quasi-symmetric since the TM$_{01}$ mode is actually a hybrid mode with a transverse electric component, in addition to other possible hybrid modes. Consequently, for a linear collider, it may be desirable to use an azimuthally symmetric structure, such as the Bragg waveguide of Fig. \ref{structures}(c), or to suppress dipole modes by careful structure design as discussed in Ref. \cite{lin:2001} for the honeycomb photonic crystal geometry of Fig. \ref{structures}(b).

\textbf{Emittance.} As a figure of merit we keep in mind that to remove the beam sufficiently from the structure's wall we assumed that $\sigma_\text{r} \leq \lambda/10$. For a heuristic estimate of the geometric emittance we note that the transverse velocity should be smaller than the velocity required for an electron on axis at the input to hit the wall $r = R$ at the exit of one acceleration stage $z = L$. Therefore, the transverse emittance must be at least of the order $\epsilon_\text{rms} = R \sigma_\text{r} / L \simeq 0.5$ nm.  For the parameter tables of Section \ref{sec:parameters} we assume a normalized emittance of $\epsilon_N = \gamma \epsilon_\text{rms}$ = 0.1 nm at $\lambda$ = 2 $\mu$m. Since $\epsilon_N$ is preserved as $\gamma$ increases, we require that $\epsilon_N$ be matched at low energy ($\gamma \simeq 1$).

\textbf{Tolerances.} The Dielectric Laser Accelerator Workshop held at SLAC National Accelerator Laboratory in 2011 examined required tolerances for a laser-driven structure based collider \cite{dla:2011}. Since the emittance must be preserved through several kilometers of acceleration, misalignments must be small enough that they do not result in significant emittance growth. Conventional magnetic focusing would require tolerances of order 1 $\mu$m in quadrupole magnet positioning, 100 nm in the accelerator structure alignment, and quadrupole jitter of less than 0.1 nm. This was based on requiring a maximum centroid motion of 10\% of the beam size from magnetic center vibration, assuming 1000 quads and a normalized transverse emittance of 0.1 nm. However, proposed electromagnetic focusing schemes which are now being incorporated into structure designs and experiments, such as alternating phase focusing and nonresonant harmonic focusing \cite{naranjo_stable_2012,niedermayer:focusing:2018}, can be built into the DLA structure design with nanometric precision that should well exceed such tolerances. A preliminary study of beam breakup instability (BBU) using a simple two-particle model found that a 30 nm average misalignment resulted in a transverse normalized emittance growth of 2.2 nm from a cold beam over 500 GeV of acceleration in 1 kilometer. A scan of emittance growth vs. bunch charge was conducted, and it was found that accelerating sufficient charge with tolerable beam degradation for high-energy physics applications requires about 50 nm alignment. Beam stability may be improved by using a shorter focusing period or by use of Balakin-Novokhatsky-Smirnov (BNS) damping. While achieving such tolerances over several kilometers is challenging, the high repetition rate of a LSA collider provides information at MHz frequencies, which can be used for feedback stabilization. Stabilization of optical components to better than 1 nm Hz$^{-1/2}$ has already been demonstrated over similar length scales at the LIGO facility \cite{ligo:stability}. Furthermore, since the acceleration process of DLA is linear with the electric field, the optical phase must be well controlled. Frequency comb technologies can detect and control both the repetition rate of the delivered pulses and the carrier to envelope phase (CEP). The technology used to generate frequency combs in ultra-high finesse Fabry Perot cavities is able to control phase noise in the range of 0.01 Hz to 100 kHz. Further stabilization will necessitate control systems operating above 100 kHz and requires important efforts in feedback loop electronics and ultrafast low-noise detectors. 

\section{Development Path}
\label{sec:development}
{\hskip 0.13in}

As part of the ANAR and ALEGRO advanced accelerator workshops \cite{anar:2017,alegro:2019}, the DLA Working Group evaluated current state of the art in the field and identified both significant advantages as well as technical challenges of the approach as a future collider technology. Key advantages include the fact that the acceleration occurs in vacuum within a fixed electromagnetic device, that the acceleration mechanism works equally well for both electrons and positrons, and that the approach is readily amenable to nanometrically precise alignment and optical stabilization of multiple stages. Complex integrated photonic systems have been shown to provide phase-stable operation for time periods of order days \cite{hulme:2014,xiang:2016}, and nanometric alignment of optical components over kilometer-scale distances has been well established by the LIGO project with stability of 0.1 nm Hz$^{-1/2}$ \cite{ligo:stability}.  In addition, the low-charge and high-repetition-rate particle bunch format inherent to this scheme would provide a very clean crossing at the interaction point of a multi-TeV collider, with estimated beamstrahlung losses in the single percent range, as compared with 10s of percents for more conventional accelerators \cite{beambeam:2021}. 

Technical challenges were prioritized from High to Low, with higher priority items being addressed earlier in the timeline. A condensed summary of corresponding R\&D topics and prioritization is presented in Table~\ref{goals}. Supporting technologies, including high average power solid state lasers and precise nanofabrication methods are lower priority, since the current state of the art in these areas is already at or near required specifications. Detailed considerations of the final focus design and beam collimation were also deemed lower priority. Due to the very low charge and low emittance beams that a LSA accelerator would intrinsically provide, existing approaches for more conventional accelerators would already be over-engineered for the LSA scenario and could thus be directly applied or perhaps even made more compact. The highest priority challenges largely pertain to the transport of high average beam currents in the relatively narrow (micron-scale) apertures of nanostructured devices.  These include effects such as beam breakup instability, charging, radiation damage, and beam halo formation, which may be less relevant at low beam powers and beam energies, but can be highly detrimental in a collider scenario.  To these ends, it was recommended to establish a core working group that would oversee the strawman collider design and motivate these feasibility studies.  

\begin{table}[]
\caption{Summary of Prioritized Development Goals}
\label{goals}
\begin{center}
\begin{tabular}{ll}
\hline
R\&D Development Area &  Priority   \\ \hline
Beam confinement, focusing, and long-distance transport   &  High \\
Longitudinal and transverse wakefield mitigation strategies & High \\
Halo, beam collimation, and beam breakup (BBU) effects & High \\
Combining of multiple parallel beams & Medium \\
Wavelength, laser requirements, and high-field damage mechanisms & Low \\
Electron and positron sources & Med-High \\
Laser-to-dielectric and-field-to particle coupling efficiencies & Medium \\
Cost drivers, power requirements, and achievable beam power & Medium \\
\hline
\end{tabular}
\end{center}
\end{table}

\subsection{Main Challenges (5 Year Time Scale)}
{\hskip 0.13in}
The LSA approach offers a number of potential benefits similar to modern radio-frequency (RF) accelerators, including the fact that the acceleration mechanism is inherently linear with incident electric field intensity and scales linearly with interaction distance, and that the particles are accelerated inside a static vacuum channel.  However, the shift in wavelength from the microwave to near infrared regime poses significant challenges for particle transport, as the beam aperture is commensurate with the wavelength of the laser (1 to 10~$\mu$m). Beam confinement in these micron-scale channels over extended distances at relativistic particle energies requires focusing fields equivalent to magnetostatic gradients of the order $10^6$ T/m, which lies beyond the capabilities of conventional techniques such as permanent magnet quadrupoles, solenoids, or electrostatic lenses.  Laser-driven electromagnetic focusing concepts have been proposed which are estimated to meet or exceed these requirements, but proof-of-principle demonstrations should be conducted as soon as possible to validate them, as well as electromagnetic modeling and particle tracking studies of focusing dynamics over relevant length scales and at  GeV and TeV particle energies.  Preliminary examination of beam breakup (BBU) in a DLA using a simple analytical model, conducted as part of a 2011 ICFA workshop at SLAC, found that maintaining beam transport with tolerable degradation for HEP applications required a beam alignment precision of approximately 50 nm \cite{dla:2011}.  More extensive numerical studies are advised to evaluate mitigation strategies, such as use of BNS damping.  Also of concern for a linear collider are the radiation damage and beam dynamics issues arising from the confinement of megawatts of average beam power.  In electron acceleration experiments conducted to date with particle energies ranging from 10 keV to 60 MeV, no particle-induced radiation damage has been observed.  However, these experiments correspond to average beam powers of less than 1 Watt.  The recent availability of the SINBAD and SwissFEL facilities at DESY and PSI, where dedicated test setups are currently under development, will enable radiation tests with 100 MeV and 3 GeV electrons. Dedicated particle tracking studies are advised to understand beam halo and beam loss effects in order to make realistic predictions of radiation dose to DLA structures at high beam power. Furthermore, the presence of MW beam power in a micron-scale aperture accelerator could lead to deleterious beam halo from dark current, transition radiation, and nonlinear space charge effects.  Many of these effects can be studied using existing codes, but others such as modeling of halo formation may require new code development. 

\subsection{Main Challenges (10 Year Time Scale)}
{\hskip 0.13in}
Technical challenges to be addressed within 10 years include those whose risk is at least partially mitigated by prior or recent progress, including: (1) active feedback for sub-micron alignment, (2) effects of high-field intensity in waveguides and materials, (3) development of high-brightness electron sources, and (4) wallplug efficiency and facility power requirements.  In order to extend the interaction over multiple wafers co-aligned with sub-micron precision, interferometric-based active feedback is necessary. Progress in this area has been made at LIGO, where intereferometric co-alignment of optical elements has been achieved with $1 \  \text{nm} / \sqrt{\text{Hz}}$ stabilization over km-scale distances \cite{ligo:2009}.  Experiments are advised to demonstrate similarly stable co-alignment of successive wafers with nanometric precision.  Additionally, due to the high (GV/m) fields necessary to drive a LSA with an appreciable accelerating gradient, dispersion and nonlinear optical effects in both the accelerators and the waveguides feeding them must be understood and mitigated in structure design.  One possible solution is to utilize hollow core waveguides to minimize high-field propagation in the material.  Prior calculations have estimated heat generation at full beam power and repetition rate to be at the level of 1 W/cm$^2$ \cite{england:rmp2014}.  Although this is well below the 1.5 kW/cm$^2$ upper limit for efficient dissipation from planar surfaces, more detailed engineering solutions should be developed, including developing photonic waveguide designs to compensate local heating issues. The restrictive acceptance of typical structure geometries implies that high brightness cathodes are necessary to generate appreciable accelerated currents. Currently, Tungsten and LaB$_6$ needle tip cathodes are promising insofar as they provide brightnesses on the order of $1 \times 10^{12}$ A/cm$^2$/sr. However, the long term stability of such cathodes is lacking. On the other hand, high brightness RF superconducting guns can potentially provide emittances on the order of nm-rad, with bunch charges in the tens of fC, and repetition rates on the MHz scale. A selection of the source to be used will be expected within 10 years.   A further means of increasing the luminosity at the interaction point is the potential use of parallelized beams. Further investigation into the emittance scaling of the recombined beam at the IP needs to be performed, however.  As a modular linac requires each laser to power multiple stages of acceleration, a network of on-chip waveguides and beam splitters is needed. Achieving critical coupling (near 100\%) of these waveguide networks is required to match the wall-plug power constraints for a linear collider facility. Although numerical studies have shown up to 95\% coupling efficiencies are possible \cite{wu:2012}, experimental validation of these results is needed.  Further, to obtain high coupling efficiency of the laser field to the electrons in the accelerating channel, optimal beam loading is required, using optically microbunched trains of particles. Theoretical studies predict efficiencies exceeding 40\% \cite{na:2005}, but over a 10 year time scale this must also be demonstrated. The ATF-II could potentially provide a testbed for such a demonstration, with their IFEL setup for CO$_2$ driven microbunching.

\subsection{Long Term View}
{\hskip 0.13in}
The required luminosity scaled from the ILC to a 3 TeV e+/e- collider is about 6$\times 10^{35}$ cm$^{-2}$ sec$^{-1}$, which corresponds to a main beam power of about 30 MW. A credible design for such a multi-TeV collider would be limited to a few hundred MW of site power, implying the site efficiency must be at least 10\% and ideally approaching 20\%. Meeting these requirements will require an integrated and well-thought out design. An international working group should be formed now to develop and maintain a strawman multi-TeV linear collider design, incorporating technology advances as they occur and identifying key remaining technology gaps. This strawman design effort would intensify after technical issues associated with small-scale optimization (i.e. focusing fields, longitudinal and transverse wakefields, alignment, and thermal effects in the dielectrics) have been addressed with short (0 to 5 years) and near term (5 to 10 years) research and once stand-alone, operating test facilities have been demonstrated. This strawman design would dictate the direction of longer term research needed to establish the validity of a collider design, with the development of a multi-stage prototype as its keystone activity. This multi-stage prototype would ideally have a dual purpose as a driver for a science application such as a future light source for cost sharing. This prototype, built within 20 years, would mimic a full linear collider design in terms of laser efficiency, power, and control; dielectric structure design; and average beam current; and would serve as a linear collider proof-of-principle demonstration motivating the construction of the full, multi-TeV LSA-based linear collider. This prototype would specifically demonstrate the feasibility of staging separate sections, extending the staging demonstration done by the STELLA-II inverse free-electron laser experiment at BNL's ATF in the early 2000s, and show that a practical system could be built without excessive beam current interception in the micron-sized apertures. It is important to note that by this stage of development, dedicated funding for a future laser-driven multi-TeV linear collider would be necessary as the goals of this stage of development would have diverged significantly from other applications. 

\subsection{Needed Technology Development}
{\hskip 0.13in}
The main supporting technologies for laser-driven structure accelerators are nanofabrication and infrared lasers, both of which have benefitted from industry-driven exponential advances over the last three decades.  Collider requirements for both of these supporting technologies were evaluated in detail as part of the Dielectric Laser Accelerator Workshop held at SLAC in 2011 \cite{dla:2011}.  Modern MEMS and CMOS fabrication capabilities exceed the spatial feature sizes of interest for LSA structure fabrication, which are on the few hundred nanometer to few micron scale, with nanometric alignment of adjacent features.  Projected laser parameters (tens of $\mu$J pulse energies, 10 to 100 MHz repetition rates, ps to sub-ps pulse duration, 40\% wallplug to photon efficiency, and of order 100 W to 1 kW average power) are at or near specifications of current solid-state Thulium, Holmium, and Ytterbium fiber laser architectures entering the market.  Furthermore, diode-pumped lasers have robust and compact form factors and have shown an exponential decrease in cost per watt combined with exponential increase in average power over the last three decades \cite{martinsen:2007}.  Consequently, the main  advances needed beyond current state-of-the-art are related to implementation:  development of efficient and robust accelerator designs and photonic waveguide networks that can be readily transferred to commercial fabrication houses for mass production; and development of feedback loop electronics and ultralow noise detectors to enable sub-cycle phase and timing control of many carrier envelope phase-locked lasers. We note that unlike the base technology development, these implementation concerns are specific to LSA and will therefore need to be independently addressed.

\subsection{Time and Cost Scales}
{\hskip 0.13in}
Progress towards an energy scalable architecture based upon laser acceleration in photonic structures requires an R\&D focus on fabrication and structure evaluation to optimize existing and proposed concepts, and development of low-charge high-repetition-rate particle sources that can be used to demonstrate performance over many stages of acceleration. To tackle these challenges, a concerted effort is required that leverages industrial fabrication capabilities and that draws upon world-class expertise in multiple areas. A number of university, national laboratory, and industrial institutions and collaborations are now actively conducting research in this area, including the multi-institutional Accelerator on a Chip International Program (ACHIP), which includes 6 universities, 1 company, and 3 national laboratories, as well as Los Alamos National Laboratory, University of Tokyo, Tel-Aviv University, The Technion, and University of Liverpool. The LSA effort could be made technology-limited through appropriate leveraging of the semiconductor and laser R\&D industries and appropriate growth of the research community. Current collaborative efforts in the U.S. and Europe are aimed at developing a first R\&D demonstration system incorporating multiple stages of acceleration, efficient guided wave systems, high repetition rate solid state laser systems, and component integration. 

Testing of speed-of-light prototype devices and initial staging experiments will require suitable test facilities equipped with relativistic beams. Existing conventional RF facilities are suitable for near-term tests over the next few years.  To provide an estimate of projected facilities costs, the current ACHIP program includes roughly \$1.3M/year of combined in-kind support from three national laboratories (SLAC, DESY, and PSI) in the form of access to personnel, resources, and beam time. However, for demonstrating many-staged on-chip accelerators, ultra-low emittance particle sources need to be developed and combined with the accelerator structures to make compatible injectors. Ongoing development of prototype integrated systems will provide a pathway for scaling of this technology to MeV, GeV, and then TeV energies and to beam brightnesses of interest both for high energy physics and for a host of other applications, as discussed in Ref. \cite{england:review:2016}.  

\section{Synergy with Other Concepts and Facilities}
{\hskip 0.13in}
\label{sec:facilities}
{\hskip 0.13in}
Laser-driven structure-based accelerators have a lot of ground to cover to be competitive with other advanced accelerator techniques which have already demonstrated gradients in excess of 50 GV/m and multi-GeV energy gains. The current state-of-the-art includes accelerating gradients approaching 1 GV/m \cite{wootton_demonstration_2016, cesar_nonlinear_2018} and energy gains on the order of 300 keV \cite{peralta:2013,cesar_pft_2018}. This approach has the potential to provide high efficiency accelerators operating at very high repetition rates, with bunch formats (charge, beam sizes, emittances, time duration and temporal separation) significantly different than what is commonly available in conventional accelerator facilities. 

For these reasons, it is important for a beam test facility to match the unique characteristics of these accelerators. For example, very high repetition rate electron sources combined with compact, efficient, and relatively low energy ($\mu$J to mJ class) lasers will enable testing of linear-collider relevant concepts such as beam loading and wall-plug efficiency. In this regard there is a general trend towards longer wavelengths, motivated in part by a desire to increase the phase space acceptance of the accelerator, which scales directly with wavelength in longitudinal and transverse dimensions. Therefore availability of suitable laser driver pulses in the mid-infrared will be needed. 

Compatibility with LSA's unique features requires high brightness beam lines equipped with diagnostics suitable for the measurement of ultralow (sub-pC to few-fC) bunch charges and ultralow (< 1 nm) normalized emittance. Recent relativistic-energy demonstration experiments have mostly taken place at low repetition rate facilities such as the Next Linear Collider Test Accelerator (NLCTA) at SLAC and the Pegasus facility at UCLA \cite{peralta:2013,wootton_demonstration_2016,cesar_nonlinear_2018}. Such facilities are sufficient for initial proof-of-principle experiments, but should in future be coupled with the novel electron sources being developed for high repetition rate free-electron lasers such as superconducting and very high-frequency RF guns. We list in Table \ref{facilities} the major current and future facilities for laser-driven acceleration experiments and the most relevant operating parameters.

\begin{table}[]
\caption{Current and Potential Test Facilities}
\label{facilities}
\begin{center}
\begin{tabular}{lllll}
\hline
Facility                & Energy & Rep Rate &  Charge   & R\&D Goals  \\ \hline
UCLA Pegasus            & 5-10 MeV   & 5 Hz  & 1 pC    &  Relativistic acceleration, phase control \\
DESY Sinbad              & 100 MeV & 10 Hz  & 1 pC   &  Relativistic microbunching, net acceleration  \\ 
Stanford         & 100 keV  & 1 MHz  & 1-1000 e     &   Tabletop MeV probes for UED, atto-science  \\ 
FAU               & 30 keV & 1 kHz     & 1-1000 e       &  Tabletop MeV probes for UED, atto-science    \\
PSI     & 3 GeV          & 100 Hz & 10 pC  &  Damage limits, diagnostics, wake effects      \\
APEX    & 750 keV  & 1 MHz    &  10 fC   & Intermediate energy, high rep rate beam studies  \\
AREAL  \cite{TSAKANOV2016284}  & 5-10 MeV  & 1-50 Hz$^a$  & 10 pC   & Relativistic microbunching    \\  \hline
\end{tabular}
\end{center}
\footnotesize{$^a$ plan to be upgraded to 1 MHz and investigation of CW operation}
\end{table}

A significant body of exploratory research has also been carried out using refurbished electron microscope columns \cite{breuer:2013,leedle:2015,leedle_dielectric_2015,mcneur_elements_2018}.  However, it should be noted that the brightness of these sources is not sufficient to efficiently couple the beam to the small phase space acceptances typical of LSA devices. Furthermore, an analysis of coupling efficiency shows that at least mildly relativistic electron energies are strongly favored to maximize the accelerating gradients in the first stages of acceleration.  Another very important characteristic will be the availability of a complete suite of optical diagnostics which should be employed whenever possible to monitor performance and provide active feedback of the laser illumination of the structures.  With all these elements available, the logical progression of experiments to demonstrate suitability for high energy physics applications includes the demonstration of bunching, beam control over longer distances, staging, wakefield mitigation, beam halo, emittance preservation, and efficient energy transfer.

It is likely that the path to a multi-TeV linear collider will include a multi-stage GeV-scale prototype to demonstrate the feasibility of the candidate collider technology or technologies to confirm gradient, emittance control, and wall-plug to beam efficiency, and to validate the fabrication cost model \cite{P5:2014}. Prior to building a large facility based upon a cutting-edge concept, a demonstration system of intermediate scale is well advised.  The primary purpose of such a demonstration system would be to incorporate interrelated technologies developed under a prior sequence of R\&D steps in order to identify and address new challenges arising from the integration of these components. Ideally, such a demonstration system would combine all or most of the technological sub-units needed to build a larger-scale system, and would simultaneously possess utility in its own right as a compelling scientific tool. We envision such a demonstration system based to consist of a sequence of wafer-scale modules, which would each incorporate of order tens of single-stage acceleration sections individually driven by on-chip fiber or SOI type guided wave systems for directing laser light and phasing them in sync with the passing speed-of-light particle beam, as described in Refs. \cite{england:rmp2014,colby:2011,dla:2011,snowmass:2013}. Such a system would illustrate:  (1) integration of the DLA concept with compatible MeV particle sources with nanometric beam emittance and attosecond particle bunch durations, (2) implementation of an accelerator architecture with a pathway to TeV beam energies, (3) carrier envelope phase-lock synchronization of multiple lasers and correct phasing and delivery of laser light over multiple acceleration stages, (4) beam alignment and steering between wafer-scale modules using interferometric techniques combined with feedback, and (5) efficient power handling and heat dissipation.  Due to the high potential cost of such a multi-stage prototype, cost sharing may be possible if the prototype also serves a secondary purpose, such as driving a future light source.

\section{Conclusion}
\label{sec:conclusion}
{\hskip 0.13in}
Particle acceleration in microstructures driven by ultrafast solid state lasers is a rapidly evolving area of advanced accelerator research, leading to a variety of concepts based on planar-symmetric dielectric gratings, hollow core fibers, photonic crystals, and plasmonic meta-surfaces. This approach, which we have referred to here as a \textit{laser-driven structure-based accelerator} (LSA), leverages well-established industrial fabrication capabilities and the commercial availability of tabletop lasers to reduce cost, with demonstrated axial accelerating fields in the GV/m range. Wide-ranging international efforts have significantly improved understanding of gradient limits, structure design, particle focusing and transport, staging, and development of compatible low-emittance electron sources. With a near-term focus on low-current MeV-scale applications for compact scientific and medical instruments, as well as novel diagnostics capabilities, structure-based laser-driven accelerators have several key benefits that warrant consideration for future high-energy physics machines, including low beamstrahlung energy loss, modest power requirements, stability, and readiness of supporting technologies.


\bibliographystyle{unsrt}
\bibliography{LSA_Snowmass2022}

\end{document}